\documentclass{iau} 
\usepackage{graphicx}
\usepackage{physics}
\usepackage{amsmath, esint}

\DeclareMathOperator{\ra}{\mathrm{Ra}}

\DeclareMathOperator{\ek}{\mathrm{E}}
\DeclareMathOperator{\pr}{\mathrm{Pr}}
\DeclareMathOperator{\prmag}{\mathrm{Pm}}
\DeclareMathOperator{\di}{\mathrm{Di}}

\title[Dipolar stability in spherical simulations] 
{Dipolar stability in spherical simulations: the impact of an inner stable zone}

\author[Zaire \& Jouve]   
{Bonnie Zaire
 \and Laurène Jouve}

\affiliation{IRAP, Université de Toulouse, CNRS / UMR 5277, CNES, UPS, 14 avenue E. Belin, Toulouse, F-31400 France
 \\[\affilskip]
$^1$email: {\tt bzaire@irap.omp.eu}}

\pubyear{2019}
\volume{xxx}  
\jname{Title of your IAU Symposium}
\editors{A.C. Editor, B.D. Editor \& C.E. Editor, eds.}
\begin{document}

\maketitle

\begin{abstract} Magnetic fields vary in complexity for different stars. The stability of dipolar magnetic fields is known to depend on different quantities, e.g., the stellar rotation, the stratification, and the intensity of convective motions. Here, we study the dipolar stability in a system with an inner stable zone. We present preliminary results of dynamo simulations using the Rayleigh number as a control parameter. The stiffness of the stable zone is accordingly varied to keep a constant ratio of the Brunt-V\"{a}is\"{a}l\"{a} frequency to the angular velocity. Similarly to the completely convective spherical shell, we find that a transition exists between a regime where the magnetic field is dipolar to a multipolar regime when the Rossby number is increased. The value of the Rossby number at the transition is very close to the one of the fully convective case.

\end{abstract}

\firstsection 
\section{Introduction}
Fully convective simulations show that the stability of dipolar dynamos depends on the Rossby number (\cite[Gastine et al. 2012]{GDW12}).  Two regimes exist: one for low Rossby number, which results in strong dipolar fields, and another with higher Rossby numbers that lead to complex magnetic field configurations. It is not clear if the same trend exists in the presence of an inner stable zone, a configuration that describes stars with a radiative core. 

Helioseismology analysis reveals the presence of a shear layer at the interface of the stable zone with the convective zone in the Sun. This layer is known to be a source of strong toroidal fields through the so-called $\Omega-$effect. Numerical simulations possessing a stable zone indicate that complex dynamics operate (\cite[Guerrero et al. 2016]{GGK16}). Although the shear is expected to impact the magnetic field generation, no parametric study using the Rayleigh number to force the intensity of the convection has been carried out for this setup using 3D simulations. 

In this work, we probe the influence of the inner stable zone in the magnetic field properties. We simulate fully convective spherical shells and partially convective shells (radiative zone  + convective envelope). In Section \ref{model} we describe the setup of the simulations and in Section \ref{mag_results} we present our magnetic results. 

\section{Models}\label{model}

We perform 3D MHD simulations in spherical geometry. Our computational domain covers a full shell with inner radius $r_\mathrm{i}$ and outer radius  $r_\mathrm{o}$. Two different setups are considered in this work: {\bf (i)} a fiducial fully convective setup, which consists of a spherical shell with aspect ratio $r_\mathrm{i}/r_\mathrm{o} =0.6$; and {\bf (ii)} a partially convective setup with aspect ratio $r_\mathrm{i}/r_\mathrm{o} =0.4$. The setup of the latter is essentially the convective zone of model {\bf (i)} with an additional inner stable zone. 

Next, we show the governing equations that describe our system and we define the reference states that capture the physical conditions of each setup.

\subsection{Governing equations and non-dimensional parameters} \label{gov_eqs}
We solve the MHD equations for a stratified fluid in a spherical shell that rotates with angular velocity 
$\Omega$ about the axis $\vu{e}_\mathrm{z}$. We adopt a dimensionless formulation where  $r_\mathrm{o}$ is the reference lengthscale,
the viscous diffusion time $r_\mathrm{o}^2/\nu$  is  the timescale, and $r_\mathrm{o}|\mathrm{d}{\bar{s}}/\mathrm{d}{r}|_{r_\mathrm{o}}$ 
is the entropy scale. Gravity, density, and temperature are normalised by their outer radius value.

We use the anelastic version of the code MagIC (\cite[Gastine \& Wicht 2012]{GW12}) to solve the non-dimensional equations that govern convective motions, magnetic 
field generation, and  entropy fluctuations:
\begin{equation} \label{eq:mom_conserv}
\ek\left[\pdv{\va{u}}{t}  + (\va{u} \cdot \grad) \va{u} \right] = - \grad(\frac{p}{\bar{\rho}}) + \frac{\ra\ek}{\pr} gs'\vu{e}_\mathrm{r}  - 2\vu{e}_z\cross\va{u} + \frac{1}{\prmag \bar{\rho}}(\curl \va{B} )\cross\va{B}
+ \frac{\ek}{\bar{\rho}}\div S ,
\end{equation}
\begin{equation} 
\pdv{\va{B}}{t}   = \curl( \va{u}\cross\va{B} ) -\frac{1}{\prmag}\curl(\curl\va{B}) \qq*{,}   
\end{equation}
\begin{equation}  \label{eq:energy_conserv}
\bar{\rho}\bar{T}\left[\pdv{s'}{t}  + (\va{u} \cdot \grad) s' +u_\mathrm{r}\dv{\bar{s}}{r} \right] = \frac{1}{\pr}  \div(\bar{\rho} \bar{T} \grad s')  
+ \frac{\pr\di}{\ra}Q_\nu + \frac{\pr\di}{\prmag^2\ek\ra}(\curl\va{B})^2 \qq*{,}  
\end{equation}
\begin{equation}
\div(\bar{\rho} \va{u}) = 0 \qq*{,} 
\div \va{B} = 0 \qq*{,} 
\end{equation}
where $S$ and $Q_\nu$ are, respectively,  the strain-rate tensor and viscous heating.
The equations above are expressed in terms of five dimensionless control parameters, which are: the Ekman number $(\ek)$, Rayleigh number $(\ra)$, Prandtl number $(\pr)$, magnetic Prandtl number $(\prmag)$, and dissipation number $(\di)$. These non-dimensional numbers are defined as follows:
 \begin{equation} 
\ek = \frac{\nu}{\Omega r_\mathrm{o}^2} \qq*{,} 
\ra = \frac{g_\mathrm{o} r_\mathrm{o}^4}{c_\mathrm{p}\kappa\nu} \left| \dv{\bar{s}}{r} \right|_{r_\mathrm{o}} \qq*{,} 
\pr = \frac{\nu}{\kappa}, \ \prmag = \frac{\nu}{\lambda}, \qq{and} \di = \frac{g_\mathrm{o} r_\mathrm{o}}{c_\mathrm{p} T_\mathrm{o}} \qq*{,} 
\end{equation}
where, $\nu$ is the viscosity, $\kappa$ is the thermal diffusivity, and $\lambda$ is the magnetic diffusivity. The dimensionless gravity profile adopted is $g(r) = -\frac{7.36 r}{r_\mathrm{o}} +  \frac{4.99 r^2}{r_\mathrm{o}^2} +  \frac{3.71 r_\mathrm{o}}{r} -  \frac{0.34r_\mathrm{o}^2}{r^2}$.

\subsection{Reference state}
Thermodynamical quantities in equations \ref{eq:mom_conserv} to \ref{eq:energy_conserv} were expressed in terms of a static (reference) state and fluctuations around it. This reference state, indicated by an overbar, is assumed to be an ideal gas nearly adiabatic given by 
\begin{equation} 
\frac{1}{\bar{T}}\pdv{\bar{T}}{r}  =  \epsilon_\mathrm{s} \dv{\bar{s}}{r}  - \di \alpha_\mathrm{o}g(r) \qq{and}  
\frac{1}{\bar{\rho}}\pdv{\bar{\rho}}{r}  =  \epsilon_\mathrm{s} \dv{\bar{s}}{r}  - \frac{\di \alpha_\mathrm{o}}{\Gamma}g(r),
\end{equation}
where the condition $ \epsilon_\mathrm{s} \ll 1$ is necessary to ensure that the governing equations still hold near adiabaticity. 

Here, $ \mathrm{d}\bar{s}/\mathrm{d}r $ is a prescribed non-adiabaticity that controls the radial stratification in the simulation domain. Stably-stratified regions occur whenever $ \mathrm{d}\bar{s}/\mathrm{d}r > 0 $, while negative gradients set convectively-unstable regions. In model {\bf (i)} convection is set by imposing $ \mathrm{d}\bar{s}/\mathrm{d}r  = -1$ in the entire radial domain,  i.e, $r \in (0.6,1.0)r_\mathrm{o}$. On the other hand, in model {\bf(ii)} the non-adiabaticity is given by
\[\dv{\bar{s}}{r}  = 
	\begin{cases}
         	 \left(\frac{N}{\Omega}\right)^2 \frac{\pr}{\ra\ek^2}, \ \ \  r < 0.6 r_\mathrm{o}, \\
        		-1, \ \ \ r \ge 0.6 r_\mathrm{o} ,      
	\end{cases}
\]
a profile that creates a stably-stratified layer for $ r < 0.6 r_\mathrm{o}$. The amplitude of this stable layer is a function of the non-dimensional numbers and the ratio of the Brunt-V\"{a}is\"{a}l\"{a} frequency ($N$) to the angular velocity.

In the next section we compare the magnetic configurations achieved on both setups. In the full set of simulations $\ek = 1.6 \times 10^{-5}$, $\pr = 1$, $\prmag = 5$, and $\di = 2.7$, where the choice of $\di$ sets a density contrast in the convective region equivalent to $4.8$ in both models. Furthermore, we chose to keep $N/\Omega=2$ in this first analysis of the stably-stratification impact on the magnetic field topology. 

\section{Magnetic results}\label{mag_results}

The set of simulations present here uses the Rayleigh number as a control parameter. We cover four different forcing levels: $\ra = 4.77 \times 10^7$, $6.25 \times 10^7$,  $7.81 \times 10^7$, and $1.56 \times 10^8$. Pair of simulations were run for each one of the forcing levels in order to access the impact of the stable zone in the magnetic configuration. These pairs of simulations correspond to the two different setups introduced in Section \ref{model}: {\bf (i)}  fully convective and  {\bf (ii)}  partially convective.

\begin{figure}[h!] 
\centering
\includegraphics[scale=0.25]{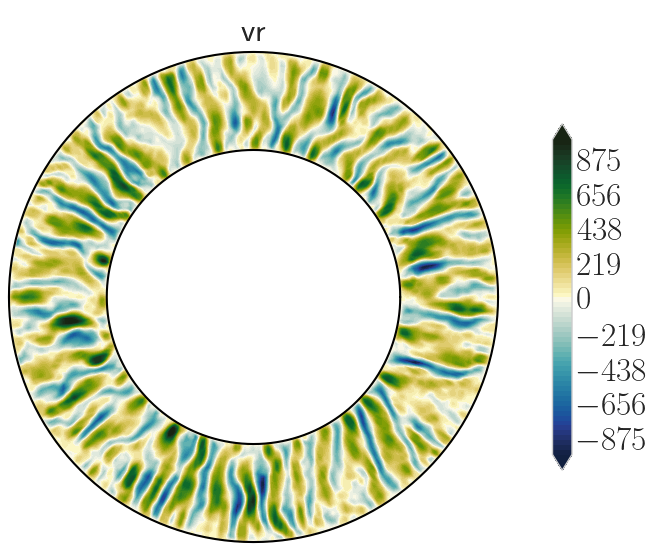}
\includegraphics[scale=0.25]{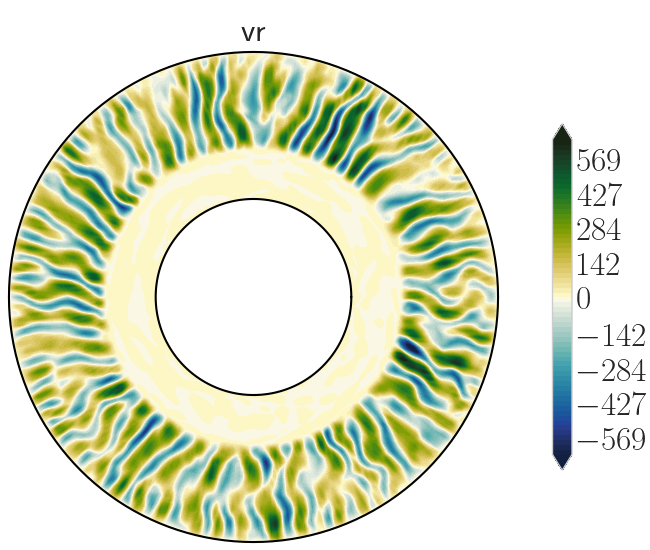}  \caption{Equatorial view of the radial velocity, $v_\mathrm{r}$, for runs with $\ra = 1.56 \times 10^8$. Left panel shows the result using the setup {\bf (i)} and right panel shows the result using the setup {\bf (ii)}. Units are the inverse of the Ekman number, $\ek^{-1}$. \label{equat}}
\end{figure}

Equatorial views of the radial velocity field are depicted in Fig. \ref{equat} for the case with $\ra = 1.56 \times 10^8$. 
Left and right panels correspond respectively to setups {\bf (i)} and {\bf (ii)}. The snapshot evidences the existence of the stable zone in setup {\bf (ii)}. Following \cite[Christensen \& Aubert (2006)]{CA06}, we define the local Rossby number as
\begin{equation*}
Ro_\mathrm{l} = \frac{u_\mathrm{rms}}{\Omega r_\mathrm{o}} \frac{\bar{l}}{\pi},
\end{equation*}
where $\bar{l}$ is the mean spherical harmonics degree in the kinetic energy spectrum. A comparison between both runs shows that the intensity of the convective flows is diminished in the presence of the stable zone, thus lowering the Rossby number. This damping behavior occurs for all partially convective simulations, however it is more prominent in the run with higher forcing.

\begin{figure}[h!] 
\centering
\includegraphics[scale=0.3]{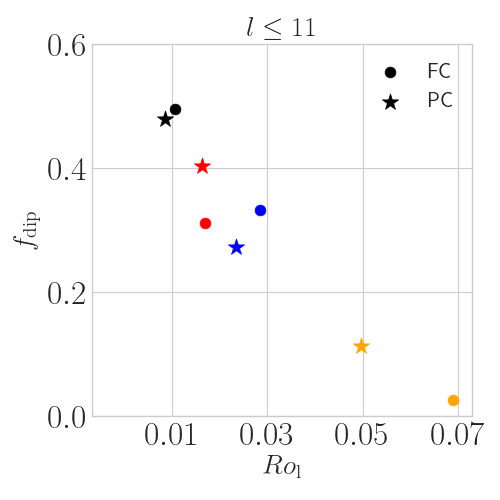}
\includegraphics[scale=0.2]{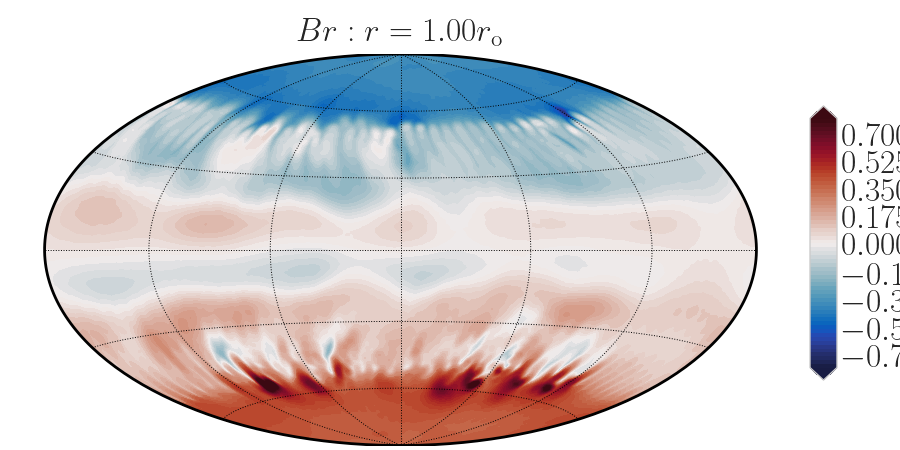}
\includegraphics[scale=0.2]{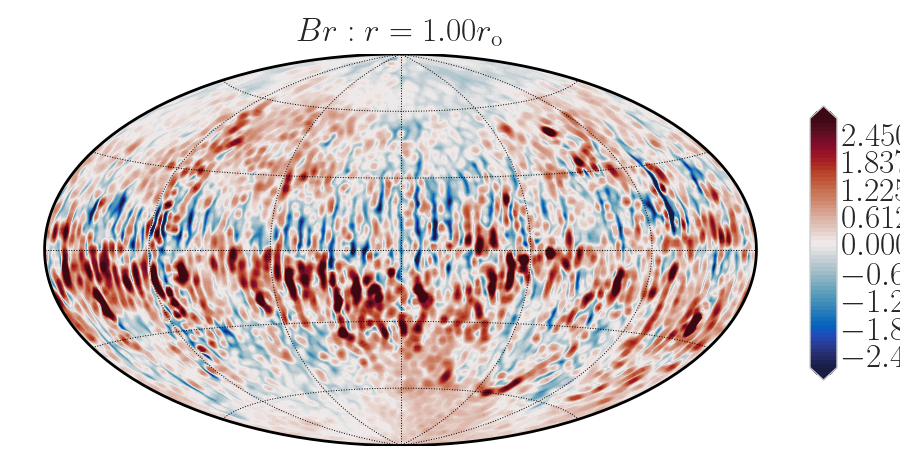}
  \caption{Left panel shows dipolar fraction in our set of simulations. Symbols denote simulations with different setups, where circles represent the setup {\bf (i)} and stars represent the setup {\bf (ii)}. Colours distinguish simulations by its forcing with black, red, blue, and orange standing for $\ra = 4.77 \times 10^7$, $6.25 \times 10^7$,  $7.81 \times 10^7$, and $1.56 \times 10^8$ respectively. Right most panels depict the magnetic topology for the partially convective setup at the lowest and the highest Rossby numbers. \label{fdip}}
\end{figure}

Figure \ref{fdip} shows the complexity of the magnetic field in our simulations as a function of the local Rossby number. We characterise the complexity of the surface magnetic field through the fraction of dipole. We express it as the relative energy in the axial dipole to the total energy in spherical harmonics degrees up to 11 at the surface:
\begin{equation}
f_\mathrm{dip}(r=r_\mathrm{o}) = \frac{\va{B}^2_{(l=1, m=0)} (r=r_\mathrm{o})}{ \sum_{l=0}^{11} \sum_{m=0}^{l} \va{B}^2_{(l,m)} (r=r_\mathrm{o})}.
\end{equation}

Magnetic fields were consistently generated in our set of simulations. The transition from dipolar to multipolar fields  was observed for both setups considered. Runs with $\ra = 4.77 \times 10^7$ (black color in Fig. \ref{fdip}), $\ra = 6.25 \times 10^7$(red) and  $\ra = 7.81 \times 10^7$ (blue) produced strong axial dipolar fields, whereas the case with $\ra =1.56 \times 10^8$ (orange) resulted in complex surface magnetic fields. For both setups the case with 
$\ra = 7.81 \times 10^7$ (blue) showed a reversal dipole, a characteristic behavior for the simulations with Rossby number at the transition for multipolar solutions.

\section{Conclusions and perspectives}

We performed numerical simulations to access the stability of dipolar solutions in the presence of a stable zone. 
Two different setups were considered, one fully convective and another partially convective. Our set of simulations corresponds to a parametric search using the Rayleigh number, which translates into a variation of Rossby number, i.e. the influence of rotation on convection.
 
For both setups dipolar and multipolar solutions were achieved. This preliminary results indicate that the dipolar transition occurs in the same region for simulations with and without stable zone. Additional runs are necessary to identify at which Rossby value the transition occurs and if other regimes exist where the dipolar solution would be favoured.


\begin{thebibliography}{}

\bibitem[Christensen \& Aubert (2006)]{CA06}
{Christensen, U. R. and Aubert, J.} 2006, 
\textit{Geophysical Journal International}, 166(1), 97-114

\bibitem[Gastine \& Wicht (2012)]{GW12}
 {{Gastine, T.} and {Wicht, J.}} 2012, 
 \textit{Icarus}, 219(1), 428-442

\bibitem[Gastine et al. (2012)]{GDW12}
 {{Gastine, T.} and {Duarte, L.} and {Wicht, J.}} 2012, 
\textit{Astronomy \& Astrophysics}, 546, A19


\bibitem[Guerrero et al. (2016)]{GGK16}
 {{Guerrero}, G. and {Smolarkiewicz}, P.~K. and
         {de Gouveia Dal Pino}, E.~M. and {Kosovichev}, A.~G. and
         {Mansour}, N.~N.} 2016, 
\textit{The Astrophysical Journal}, 819:104

\end{thebibliography}
\end{document}